# IS FORTRAN STILL RELEVANT? COMPARING FORTRAN WITH JAVA AND C++

Shahid Alam

Department of Computer Science, University of Victoria, BC, Canada

**ABSTRACT**

*This paper presents a comparative study to evaluate and compare Fortran with the two most popular programming languages Java and C++. Fortran has gone through major and minor extensions in the years 2003 and 2008. (1) How much have these extensions made Fortran comparable to Java and C++? (2) What are the differences and similarities, in supporting features like: Templates, object constructors and destructors, abstract data types and dynamic binding? These are the main questions we are trying to answer in this study. An object-oriented ray tracing application is implemented in these three languages to compare them. By using only one program we ensured there was only one set of requirements thus making the comparison homogeneous. Based on our literature survey this is the first study carried out to compare these languages by applying software metrics to the ray tracing application and comparing these results with the similarities and differences found in practice. We motivate the language implementers and compiler developers, by providing binary analysis and profiling of the application, to improve Fortran object handling and processing, and hence making it more prolific and general. This study facilitates and encourages the reader to further explore, study and use these languages more effectively and productively, especially Fortran.*

## 1. INTRODUCTION AND MOTIVATION

One of the most important aspects of a programming language is its ability to provide higher levels of abstractions to the programmer. A programmer can define abstract data types such as: types with extensions in Fortran, interfaces in Java and classes in C++.

According to the *Tiobe Language Popularity Index* [17] the three most popular programming languages in any category (including general-purpose, compiled, script and other) are Java, C and C++ in that order. Fortran is listed at number 33 in the index. Another *Language Popularity Index* [47] lists the three most popular programming languages in the *general-purpose* and *compiled category* as C, Java and C++ in that order. Fortran is listed at number 12 in the index. These indexes are based on searching the web (Google, MSN and Yahoo etc) with certain phrases that contain the language names and counting the number of hits returned. These indexes show that Fortran is not very popular as a general-purpose programming language. An article, published by ACM Queue, *The Ideal HPC Programming Language* [34], argues that Fortran is still one of the popular and primary languages in high performance computing (HPC).

As COBOL is popular in large corporate business data centers and C in embedded and operating systems, **Fortran** [16] has been in use and popular for over 50 years in scientific and engineering communities. Most of the compute intensive tasks such as weather and climate modeling, computational chemistry and physics, and others are programmed in Fortran. Most of the SPEC CPU2006 floating point benchmarks [23] are implemented in Fortran.

**Java** [31] compared to Fortran and C++, is a relatively new language and was first released in 1995. It is mostly used for application development, specifically for the web. It has a simpler object model and fewer low level facilities than C++. Applications written in Java can run on any machine with a JVM (Java virtual machine). This makes it a *write once, run anywhere language* and hence a popular choice among the software developers.

**C++** [13] is mostly used to write system and application software, high performance server and client applications, and video games. C++ is an enhancement to C and was initially called C with classes. The previous C++ standard [11] was published in 1998 and revised in 2003. The recent standard [13] known as C++11 was passed in 2011.

There is a large code base including libraries of Fortran that exist and are being developed by scientists and engineers for specialized applications. Efficient development and maintenance of this code is key to the success of the Fortran language. To improve these software development processes, Fortran has gone through some major and minor extensions in 2003 and 2008. The most recent standard known as Fortran 2008 [16] with these extensions was passed in 2010. Fortran and C++ both have been officially approved by the ISO standards committee. Java does not have an official standard approved by any of the standards committee but has achieved a dominant position by public acceptance and market forces. This paper presents a comparative study to evaluate and compare Fortran 2008 with Java and C++ making a case on behalf of the software developers from the scientific and engineering communities that Fortran is still relevant, and to highlight some of it's significant advantages. We are interested in answering the questions: (1) How much have the new extensions made Fortran comparable to Java and C++? (2) What are some of the similarities and differences in supporting features like: Templates, object constructors and destructors, abstract data types and dynamic binding?

These three languages are traditional languages and are being used for developing large software applications. The program for comparison should be large enough to cover most of the features of the programming language. A simple program will not reveal all the similarities and differences among the compared languages, and a very large program can make the comparison complex and the results unusable.

A basic *ray tracer* is implemented to compare the three languages. The *ray tracing* application implemented in this paper is neither complex nor simple, but is practical and complete enough that it has been used to generate molecular model animations (visualizations). One of them is shown in Figure 1. The *ray tracer* is an object-oriented application and can either render one image or more than one images (animation). Depending on the complexity of the animation the rendering can take a lot of CPU cycles. We also give a comparison of the runtime of the *ray tracing* application for each language and highlight some of the major differences of object handling and processing. While generating animation the *ray tracer* processes a lot of images and writes them to the disk. A correct measurement of the size and careful inspection of the code will give us an insight into the complexity, quality of the code, similarities and differences in each language.

The rest of the paper is organized as follows: Section 2 reviews some of the previous studies carried out to compare Fortran with other languages. Section 3 describes the design of the *ray tracer*. The object oriented features of the *ray tracing* application implemented in the three languages are compared and discussed using the size and class level, and basic software, metrics in Sections 4 and 5. Section 6 provides binary analysis and profiling of a simple object-oriented application to highlight and discuss, where Fortran compiler lacks in optimizations, for object handling and processing. Finally we conclude in Section 7.

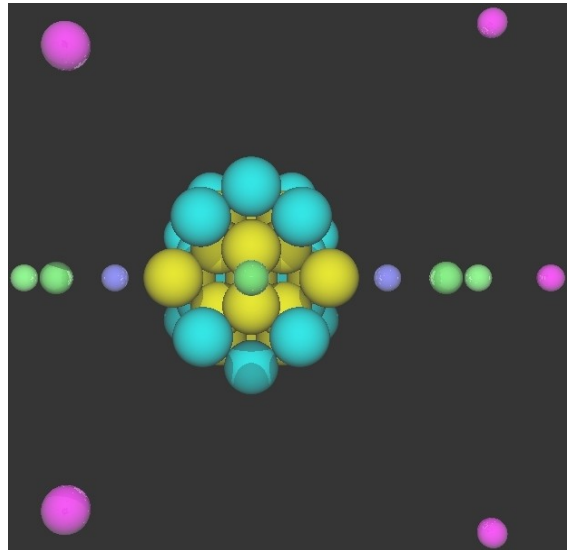

**Figure 1: Animated Chlorine Molecular Model Rendered by the *Ray Tracer*. Only the first image (frame), out of 1245 rendered images (frames), is shown here. This and other animations are available online @ [5].**

## 2. LITERATURE REVIEW

In this literature review we only cover the published literature that compares Fortran with other popular languages including C++, Java, Pascal and Simula.

The first comparative study of Fortran with Simula, an object-oriented language, was done by Jacob et al [43] in 1968. A simulation program was used for comparison. The authors found that the program in Simula was 24% shorter but the runtime was 64% longer, than Fortran. They also concluded that Fortran was more machine efficient where as Simula was more programmer efficient (productive). The next comparative study of Fortran was done with Pascal by Garry et al [42] in 1978. The purpose of this study was to describe how and why the transition was made from Fortran to Pascal, for teaching as an introductory programming language, in the department of computer science at the University of Colorado. At that time Pascal provided more features than Fortran such as, extensible data and better control, structures. Moreover Fortran could be easily learned once the student was comfortable with Pascal.

John et al [9] and Norton et al [40, 39] compares the object-oriented features of Fortran 90 [14] with C++ for scientific programming. The major new features of Fortran 90 are type, module and dynamic memory management. The authors were able to model many aspects of the object-oriented features of C++ using Fortran 90. But their conclusion was that this modeling can be tedious and inappropriate at times. Fortran 90 does not fully support object-oriented features. This lack of support of object-oriented and other features in Fortran 90 has lead to the extension of Fortran to support more features such as inheritance, polymorphism, procedure pointers and Co-arrays etc in 2003 [15] and 2008 [16]. These extensions are the basic motivations for the study carried out in our paper. The paper [9] mentioned the templates as the critical missing feature in Fortran 90. This feature was not included in the latest Fortran 2008 standard or in Java because of the cost of virtual function calls [28].

Mosli et al [37] compare Fortran 90 [14] with C++ and Oberon-2 [38]. Only the main language features are compared, such as data types, pointers and arithmetic operators etc, and not the object-oriented features. Bull et al [7] compares the performance of Java against C and Fortran for scientific applications. The authors used Java Grande benchmark suite [8] for comparison and concluded that the performance gap of Java on some platforms (Intel Pentium) compared to C and Fortran is quite small.

Moreira et al [36] in 1998 compared the performance of Fortran, C++ and Java for numerical computation, which involved a lot of array computation and manipulation. The authors discussed the support of optimizations in the compilers for each language and how to make the compilers for Java and C++ as efficient as Fortran for array processing. A matrix multiplication example was used for comparison. According to the authors Fortran demonstrated better performance than Java and C++. Based on our literature survey the study carried out in this paper is the first of its kind to evaluate and compare Fortran with C++ and Java, in practice: implementation and analysis of the *ray tracing* application, and in theory: applying software metrics to the *ray tracer*.

## 3. DESIGN OF THE RAY TRACER

Figure 2a shows a flow chart of the *ray tracing* application. The complete source code for the *ray tracer* in all the three languages is available online @ [5]. The *ray tracer* implemented here only renders spheres and planes. The rendering of images (scene) is done by ray tracing each ray from the camera. The number of rays depends on the height and width of the scene. A scene file is used to define spheres, planes, lights, materials and paths in the scene. The *ray tracer* generates animation by using the object Path defined in the scene file. A Path is attached to an object in the scene file to generate animation for that object. The Path contains information of an elliptic path. The *ray tracer* uses this information to compute the positions around the elliptic path. Then it generates images at each position. These images together create an animation. More details about the scene file, rendering of animation and the design of the *ray tracer* is given in [6]. One of the animations, visualization of a chlorine molecular model, rendered using the *ray tracer* is shown in Figure 1.

The function *RayTrace()* is compute intensive. The pseudocode of this function is listed in Figure 2b. The computation intensity depends on the size of the scene, number of objects in the scene, and the number of reflections and refractions by the ray. At each iteration the function *RayTrace()* checks all the objects in the scene for a hit by the ray. If a ray hits an object the shading, shadow, reflection and transparency/refraction are computed, which are then used to compute the color of the pixel at the hit position. The reflection and refraction is performed by recursively iterating on the reflected and refracted rays.

### 3.1. Methodology used for Implementing the Ray Tracer

In this section we describe the methodology used to implement the *ray tracing* application in the three languages.

Out of the three languages C++ is the first language that provided OOP (Object-oriented programming) support. We first implemented the *ray tracing* application in C++ as an object-oriented application. It was then ported to Java and Fortran in that order. The purpose of this study is not to compare the ease of learning or programming in the language, or comparing the application implementation by persons with different skill levels in each language. Therefore only one person was chosen to implement the application who was equally good in the three languages. That is why the results of object-oriented software metrics as shown in Tables 1 and 2are mostly similar for the three languages. This shows that Fortran provides similar object-oriented functionalities and features as the other two languages, which is one of the purposes of the paper.

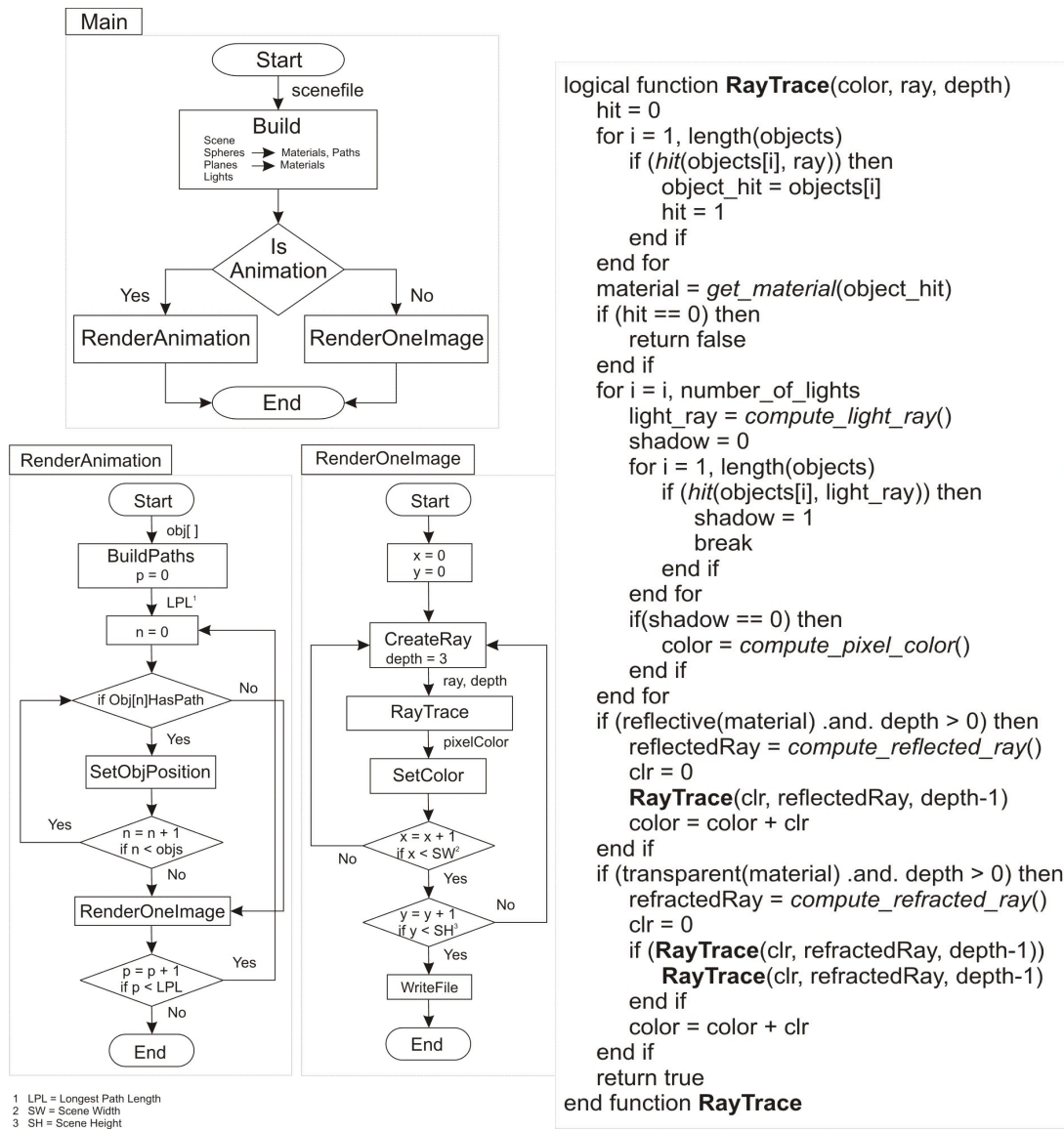

(a) *Ray Tracer* Flow Chart    (b) Pseudocode of the *RayTrace* Function

**Figure 2:** *Ray Tracer* **Flow Chart and the Function** *RayTrace*

No deliberate effort was made to optimize the source code manually in any of the three languages. Except, to make the source code as equivalent as technically possible for comparison, in Fortran we change, at various places, the indirect access of objects to direct as shown in the code snippet below:

```
From:
   do i = 1, World%Scene%GetNumberOfSpheres()
      sphere_t = World%Scene%GetSphere(i)    ! Method call to access the private object
To:
   do i = 1, World%Scene%GetNumberOfSpheres()
      sphere_t = World%Scene%sphere_t(i)     ! Accessing directly the public object
```

We believe emphasizing on one application provides more detailed analysis of the language and the compilers than the traditional micro benchmarks. We provide this insight and analysis in Sections 5 and 6. We make the case of why this application was chosen, and is suitable for comparing the three languages in Sections 1 and 5.1.

One important aspect of the *ray tracing* application that we want to highlight here is the way this application handles and processes the objects. This processing may be different than other micro benchmark applications, but focusing on this one application helped us compare this property and highlight some differences in the three object-oriented languages. This comparison and the differences are discussed in detail in Section 5.

## 3. APPLYING BASIC OBJECT-ORIENTED SOFTWARE METRICS TO THE RAY TRACER

Tables 1 and 2 show size-level and class-level metrics [2, 45] for the *ray tracing* application. We have only selected basics of these metrics to represent complexity and quality of the object-oriented features of the *ray tracing* application for comparing the three languages. The purpose of applying these metrics is to highlight the differences and similarities among these languages such as the same depth of inheritance achieved in all three languages, and that there are a greater number of classes in Fortran and Java than C++. Overall there are more similarities than differences that proves our conjecture, that Fortran as an object-oriented language is equivalent to one of the two most popular object-oriented languages Java and C++. In the next Sections we will see that there are some differences that need to be improved to make Fortran a more relevant object-oriented language.

### 4.1. Selection of Object-Oriented Software Metrics

This Section describes the selection of the object-oriented software metrics for the case study carried out in this paper. To compare the object-oriented features of Fortran with the other two languages we selected 3 size-level metrics and 5 class-level metrics. The comparison highlights both the differences and similarities in the three languages. These metrics are explained in more detail in Section 4. Here we give the reasons for selecting these metrics as follows:

**Size-Level Metrics**: An object-oriented language provides the ability to create objects (i.e classes). These objects have methods and attributes, that are used to define the behaviour and the structure of the object, respectively. We wanted to see, as an object-oriented language, how Fortran compares with the other two languages based on this basic property. So we selected 3 size-level metrics. TC, TM and TA: These are the total number of classes, methods and attributes respectively in an application implemented using an object-oriented language. **Class-Level Metrics**: An object or a class is a basic construct in an object-oriented language. This construct lets the software developer design a code that is easy to develop, maintain, reuse and test. Because of there pivotal role in an object-oriented system we are particularly interested in the following class-level basic properties of an object-oriented language.

Data abstraction: This is the ability to hide a complex (e.g: a user defined data) data in a class. Such a data with a well defined interface (easy to communicate with other classes) can be used without knowing the details of its implementation. This reduces the software development time and cost. To compare this property we selected the metric **DAC** (data abstraction coupling). The coupling here means the ability of a class to communicate with other classes. Inheritance: This provides reusability of the code and is achieved by deriving a new class from an existing (a parent) class. The new class inherits all the features of the parent class and yet more features can be added to the new class. The metric **DIT** (depth of inheritance) was selected to measure this property. This metric deals with the relationship among classes that enables programmers to reuse previously defined objects. Polymorphism: The ability to take more than one form. A method of a class can behave differently in different instances of the class. This provides easy maintenance and reusability of the code. The metric that we used in this paper to measure this property is **NMO** (number of methods overridden by a class). This metric counts the number of polymorphic methods in a class. The other two metrics **NOM** (number of methods per class) and **NOA** (number of attributes per class) are the size-level metrics at the class-level. The metrics DIT and NMO both focus on the inheritance property of a class but have different meanings.

In the next two Sections using these metrics and the source code listings in Figures 3 and 4 we will highlight some of the object-oriented feature differences and similarities in the three languages.

**Table 1: Size-Level Metrics for the *Ray Tracing* Application in Fortran, C++ and Java**

| Fortran | | | C++ | | | Java | | |
|---|---|---|---|---|---|---|---|---|
| TM | TA | TC | TM | TA | TC | TM | TA | TC |
| 151 | 78 | 19 | 165 | 82 | 17 | 154 | 101 | 24 |

### 4.2. Size-Level Metrics

There are 19 in Fortran, 17 in C++ and 24 classes in Java as shown in the Size-Level Metrics in Table 1. This difference is because there are no struct or union types in Fortran and Java. We used types in Fortran and classes in Java to imitate type struct, shown in Figure 3 as type(Color) and class Color. Also in Java the global definitions are defined in a class and the parameters for the recursive function *RayTrace()* are passed as classes. Java has more in total, but less mean number of attributes (member variables of a class) than Fortran and C++ because of the greater number of classes in Java. This confirms that Java is a pure object-oriented language and everything is an object in Java. In Fortran and C++ the programmer is not bound to use objects.

**Class Constructor**: In C++ and Java the class constructor has a much simpler declaration than Fortran. In Figure 3 the Fortran class TGA's constructor is declared as interface TGA with procedure *init_TGA()*. **Class Destructor**: As mentioned before there is no class destructor in Java and the deallocation of objects is taken care of by the garbage collector. C++ uses the symbol ~ to declare the class destructor as shown TGA::~TGA() in the Figure. In Fortran a destructor for a class is declared using the keyword final and is implemented using the procedure *final_TGA()* in the Figure. The dynamic memory allocated for the object Color in Figure 3 is deallocated, in Java by the garbage collector (controlled by the JVM - Java Virtual Machine - and not by the application), in C++ when the statement delete(tga) is called and in Fortran just before the end program example statement is reached.

As shown in the Size-Level Metrics graph in Table 1 the total number of methods in the Fortran source code of the *ray tracing* application is less than the total number of methods in the C++ and the Java source code. Because the *init* (Class constructor) and the *final* (Class destructor) methods were not implemented for all the Fortran classes. The gfortran compiler version 4.7 used in this paper supports quite a few of these Fortran 2003 [15] and 2008 [16] extensions but does not support all of them. The class constructor and destructor are not supported by the gfortran compiler as declared in Figure 3. For a complete list of Fortran compilers and their support for the Fortran 2003 and 2008 standards the reader is referred to [10].

### 4.3. Class-Level Metrics

An explanation and discussion of the Class-Level Metrics shown in Table 2 follows. The Min number for each of these metrics is 0 and the Mean is the arithmetic mean:

```
module TGA_m
  type :: Color
    integer :: red, green, blue
  end type Color
  type :: TGA
    private
      integer :: bitmap, type, rows, cols
      type(Color), allocatable :: color_t(:)
    contains
      final :: final_TGA
  end type TGA
  interface TGA
    module procedure init_TGA
  end interface TGA
  contains
  type(TGA) function init_TGA(b, t, r, c)
    integer, intent(in) :: b, t, r, c
    init_TGA%bitmap = b
    init_TGA%type = t
    init_TGA%rows = r
    init_TGA%cols = c
    allocate ( init_TGA%color_t(r*c) )
  end function init_TGA
  subroutine final_TGA(tga_t)
    type(TGA), intent(inout) :: tga_t
    if ( allocated(tga_t%color_t) ) then
      deallocate(tga_t%color_t)
    end if
  end subroutine final_TGA
end module TGA_m
! Instantiation of the Fortran TGA class
program example
  use TGA_m
  type(TGA) :: tga_t
  tga_t = TGA(24, 2, 1024, 768)
end program example
```

```
class TGA {
  private:
    struct Color {
      int red, green, blue;
    };
    Color *color;
    int bitmap, type, rows, cols
  public:
    TGA(int b, int t, int r, int c);
    ~TGA(void);
};
TGA::TGA(int b, int t, int r, int c) {
  bitmap = b;
  type = t;
  rows = r;
  cols = r;
  color = new Color[rows * cols];
}
TGA::~TGA(void) {
  if (color)
    delete color;
}
// Instantiation of the C++ TGA class
void example(void) {
  TGA *tga = new TGA(24, 2, 1024, 768);
  delete(tga);
}

public class TGA {
  private class Color {
    int red, green, blue;
  }
  private Color color[ ];
  private int bitmap, type, rows, cols
  public TGA(int b, int t, int r, int c) {
    bitmap = b;
    type = t;
    rows = r;
    cols = r;
    color = new Color[rows * cols];
  }
}
// Instantiation of the Java TGA class
public void example(void) {
  TGA tga = new TGA(24, 2, 1024, 768);
}
```

(a)  Fortran             (b) C++ and Java

**Figure 3: Part of the Source Code of the *Ray Tracer* Class TGA in Fortran, C++ and Java**

**Table 2: Class-Level Metrics for the *Ray Tracing* Application in Fortran, C++ and Java**

| Metric | Fortran | | | C++ | | | Java | | |
|---|---|---|---|---|---|---|---|---|---|
| | Min | Max | Mean | Min | Max | Mean | Min | Max | Mean |
| DAC | 0 | 8 | 2.4 | 0 | 8 | 2.6 | 0 | 10 | 2.8 |
| DIT | 0 | 1 | 0.16 | 0 | 1 | 0.18 | 0 | 1 | 0.14 |
| NMO | 0 | 3 | 0.5 | 0 | 3 | 0.53 | 0 | 3 | 0.4 |
| NOA | 0 | 28 | 4.7 | 0 | 30 | 4.8 | 0 | 29 | 4.4 |
| NOM | 0 | 26 | 8 | 0 | 27 | 9.8 | 0 | 26 | 6.7 |

International Journal of Software Engineering & Applications (IJSEA), Vol.5, No.3, May 2014

1. NOA: Number of attributes per class. An attribute represents the structural properties of a class and is defined as part of the declaration of the class. All three languages have almost the same mean number of attributes per class except Java, because of the use of classes in place of *struct*.
2. NOM: Number of methods per class. A method is an operation upon an object and is defined as part of the declaration of a class. The Mean number of methods per class in the Fortran and the Java is less than the C++. As mentioned before there are more classes in the Fortran and the Java than the C++ implementation. These extra classes are user defined data types and have 0 number of methods.

```
type, public, abstract :: Shape
  contains
    procedure(SetPosition_d), pass, deferred :: SetPosition
    procedure(GetPosition_d), pass, deferred :: GetPosition
    procedure(Hit_d), pass, deferred :: Hit
end type Shape
abstract interface
  subroutine SetPosition_d(S, X, Y, Z)
    import Shape
    class(Shape), intent(inout) :: S
    double precision, intent(in) :: X, Y, Z
  end subroutine SetPosition_d
  type(Vector) function GetPosition_d(S)
    import Shape, Vector
    class(Shape), intent(in) :: S
  end function GetPosition_d
  logical function Hit_d(S, R, t)
    import Shape, Ray
    class(Shape), intent(in) :: S
    class(Ray), intent(inout) :: R
    double precision, intent(inout) :: t
  end function Hit_d
end interface

type, public, extends(Shape) :: Sphere
  contains
    procedure, pass :: SetPosition=>SetPosition_SPHERE
    procedure, pass :: GetPosition=>GetPosition_SPHERE
    procedure, pass :: Hit=>Hit_SPHERE
end type Sphere

type, public, extends(Shape) :: Plane
  contains
    procedure, pass :: SetPosition=>SetPosition_PLANE
    procedure, pass :: GetPosition=>GetPosition_PLANE
    procedure, pass :: Hit=>Hit_PLANE
end type Plane
```

**Figure 4: Part of the Source Code of the *Ray Tracer* Classes Shape, Sphere and Plane in Fortran**

3. DIT: Depth of inheritance. This is the number of ancestor classes also called super classes to a sub class. In our *ray tracing* application there are two superclasses: *Shape* (Sphere and Plane) and *Image* (TGA). All the languages have exactly the same number of superclasses. The difference in Mean is because of the difference in total number of classes. Similar to Java, Fortran uses the keyword extends to inherit a super class. As shown in Figure 4 the super class Shape is inherited by the two sub classes Sphere and

   Plane. The inheritance in Fortran is very similar to C++ and Java. In Fortran the sub classes inherit all the attributes and methods of the super class. Out of the three languages only C++ supports multiple inheritance and Templates.

4. NMO: Number of methods overridden by a class. This makes a class polymorphic also called ad-hoc polymorphism [48]. A method defined in a super class is implemented differently in each of the sub classes. In Fortran the keywords abstract and deferred are used to declare a virtual function. In our *ray tracing* application, as shown in Figure 4, the sub classes Sphere and Plane implement the three virtual functions *SetPosition()*, *GetPosition()* and *Hit()* declared in the super class *Shape*. All of the three languages provide facilities to abstract either a complete class or an individual function. One of the super classes, Image, is an interface. Three functions in the Shape class are declared abstract in all three languages, an example in Fortran is shown in Figure 4. These functions are implemented by the sub classes Sphere and Plane. The difference in Mean is due to the difference in the total number of classes.

5. DAC: Data abstraction coupling. It is the ability to create new data types called abstract data types (ADTs). DAC is the number of ADTs defined in a class. Some of the ADTs in our *ray tracing* application are: Vector, Ray and RGBColor. Java is a pure object-oriented language and the programmer is bound to use objects (classes) for every user defined data structure. Therefore the *ray tracing* application implemented in Java has more data structures defined as classes (e.g: in the *ray tracing* application implemented in Java various *struct* types are defined as ADTs) and hence a bigger DAC number when compared to Fortran and C++.

We have discussed some of the differences and similarities in the three languages by looking at the source code and applying object-oriented metrics to the *ray tracing* application. These metrics and the discussion imply that Fortran is an equal value object-oriented language like C++ and Java. The syntax may be different but Fortran supports most of the object-oriented features present in C++ and Java. Some of the features like Templates (Fortran Parametrized Derived Types and Java Generics are technically different than Templates as listed in Section 1) and multiple inheritance are not supported in Fortran for the same reason they are not supported in Java due to their runtime overhead and the complexities of implementing an optimizing compiler. Templates with many lines of code cannot be inlined and may incur runtime overhead. For details of, why multiple inheritance make things complex and increases the runtime overhead the reader is referred to [24]. C++ and Java are more concise and clear, but Fortran is more verbose and explicit.

## 5. BASIC SOFTWARE METRICS FOR THE RAY TRACER

Figures 5 and 6 shows and Table 3 lists some of the basic software metrics for the *ray tracing* application, including the runtime and the size of the *ray tracing* application. The reason for including the size of the *ray tracing* application is to: show that Fortran is a verbose language and why it took more time to implement the application in Fortran. We have chosen four parameters to enumerate the size: number of files, number of bytes, words and lines of the source code. We believe all these parameters together give a correct measurement of the size of the code and also give an insight into the complexity of the code. The number of words and physical lines of source code also includes comments. These comments are almost the same for each language. We used the following timing functions, as recommended by all the three languages standard, in this paper to compute the CPU time: *cpu_time()* in Fortran (ref: Section 13.7.42 of [16]), *clock()* in C++ (ref: Section 7.23.2.1 of [12]) and *nanoTime()* in Java [20].

The runtime shown in Figures 5 and 6 is the CPU time in milliseconds to render each image for generating the animation. The scene files that were used to generate animations are listed in [6].

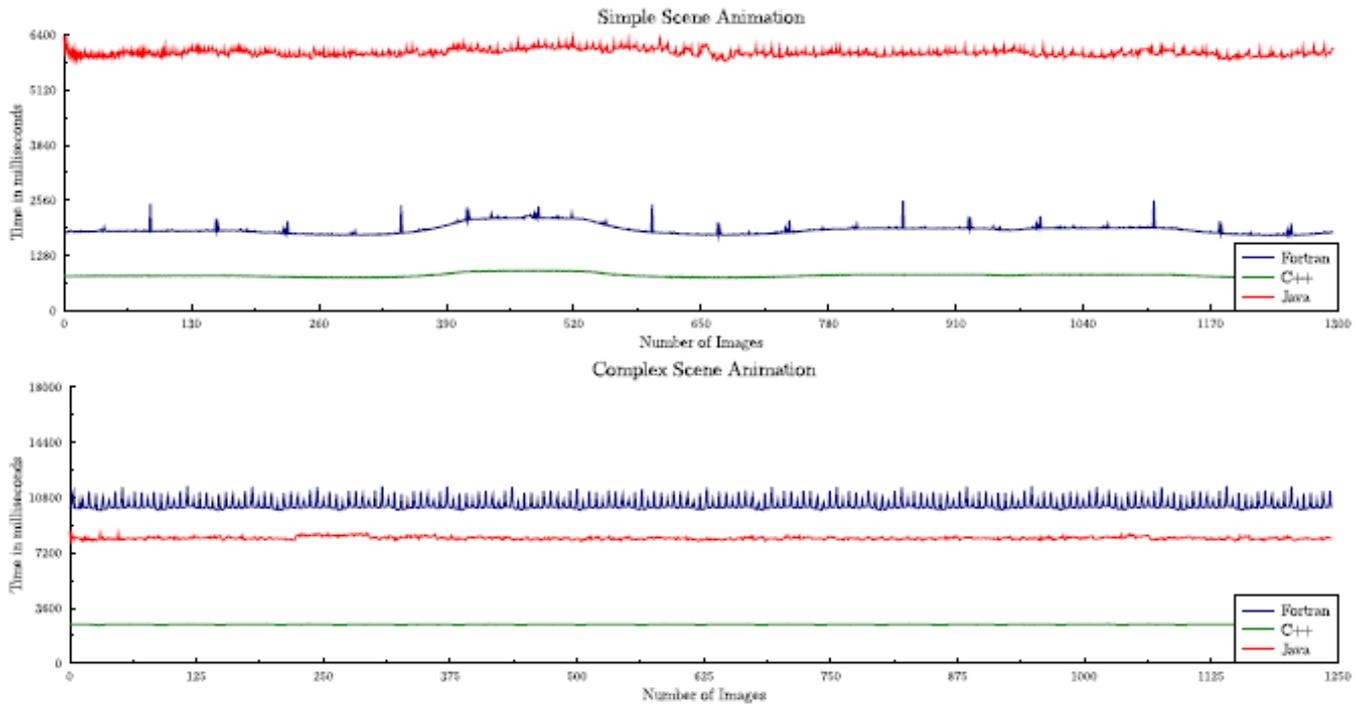

(a) Ubuntu 10.1, 64 bit, with Linux kernel 2.6.32.24

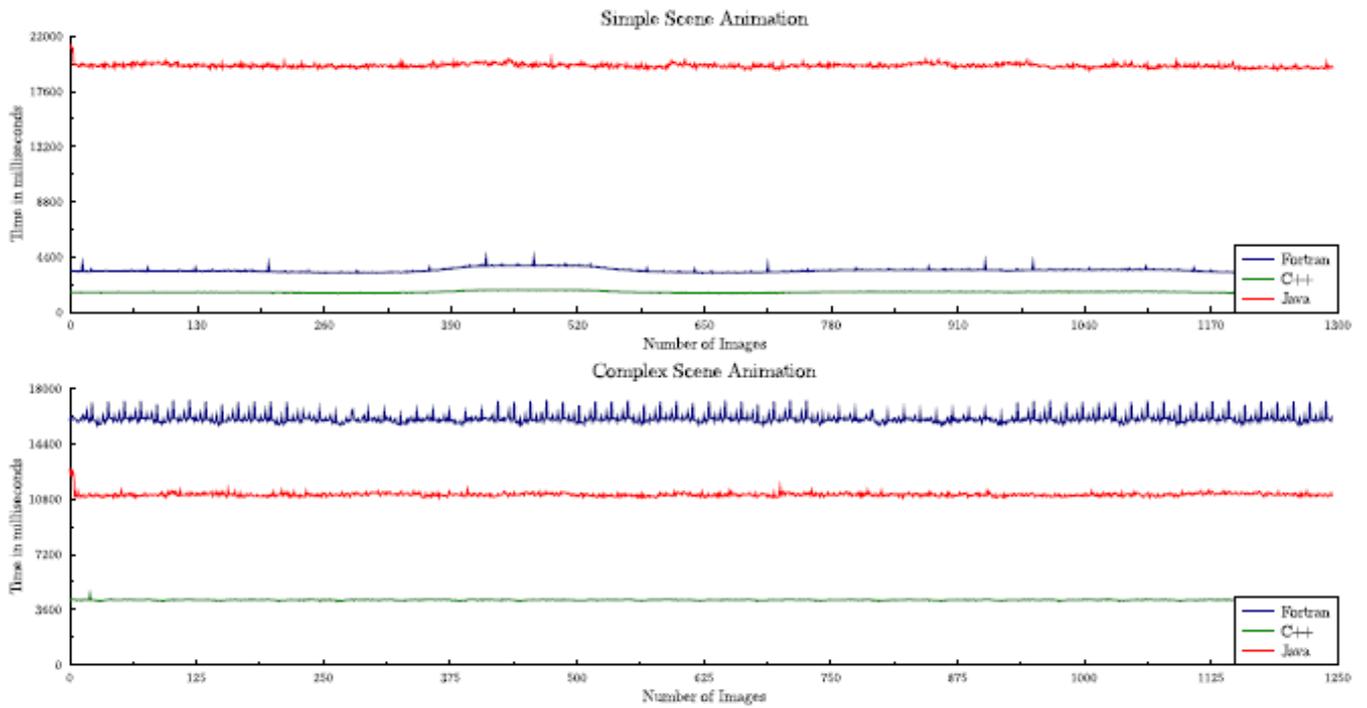

(b) Cygwin 1.7.9, 64 bit

**Machine** used: Intel Core2 Quad CPU Q6700 4GB RAM; Dedicated for the experiment i.e; why the results are more uniform than shown in Figure 6.
**Compilers**: gfortran 4.7 [25], g++ 4.7 [25] and Oracle Java 1.6.0.21 [21] **Optimization** Flags: -O3 for gfortran and g++, and no flag used for Java.

**Figure 5: Runtime of the *Ray Tracing* Application (Using GNU Compilers), in Fortran, C++ and Java, for Rendering Images to Generate Animation.**

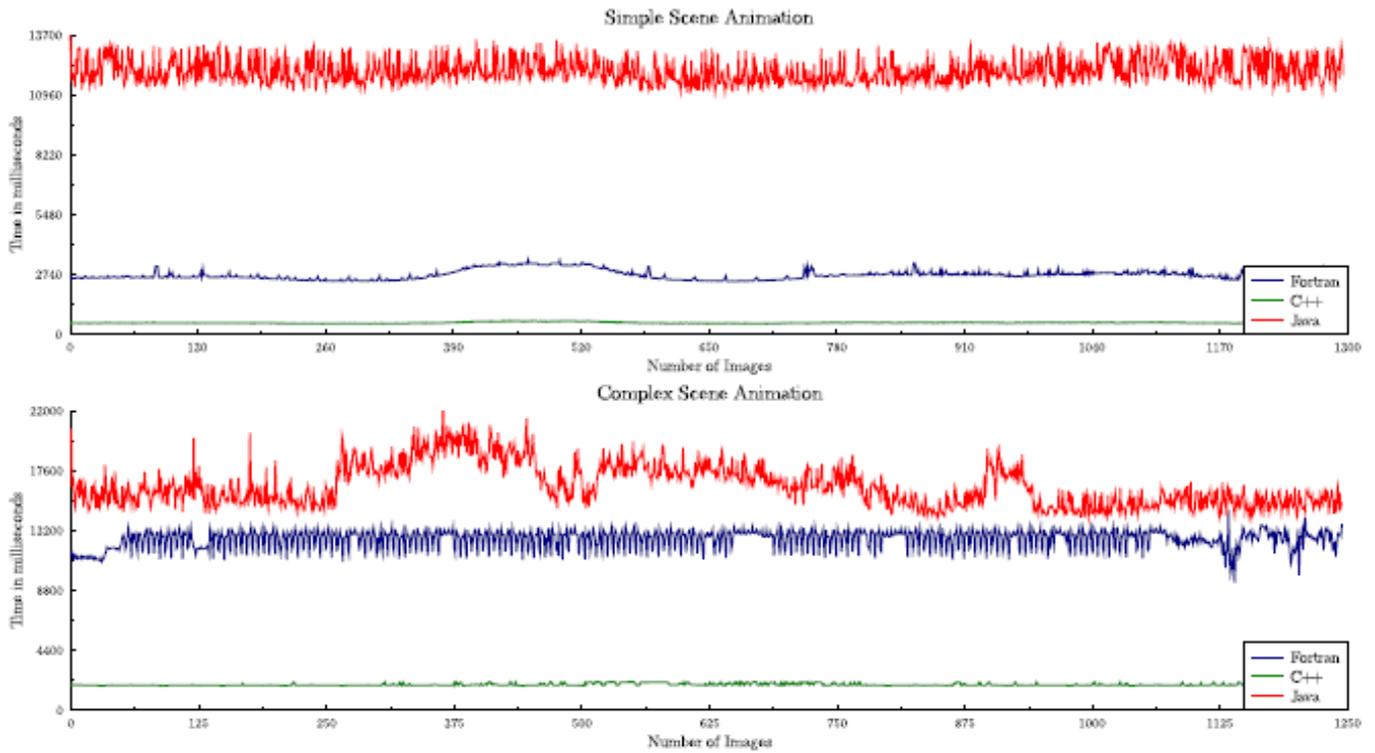

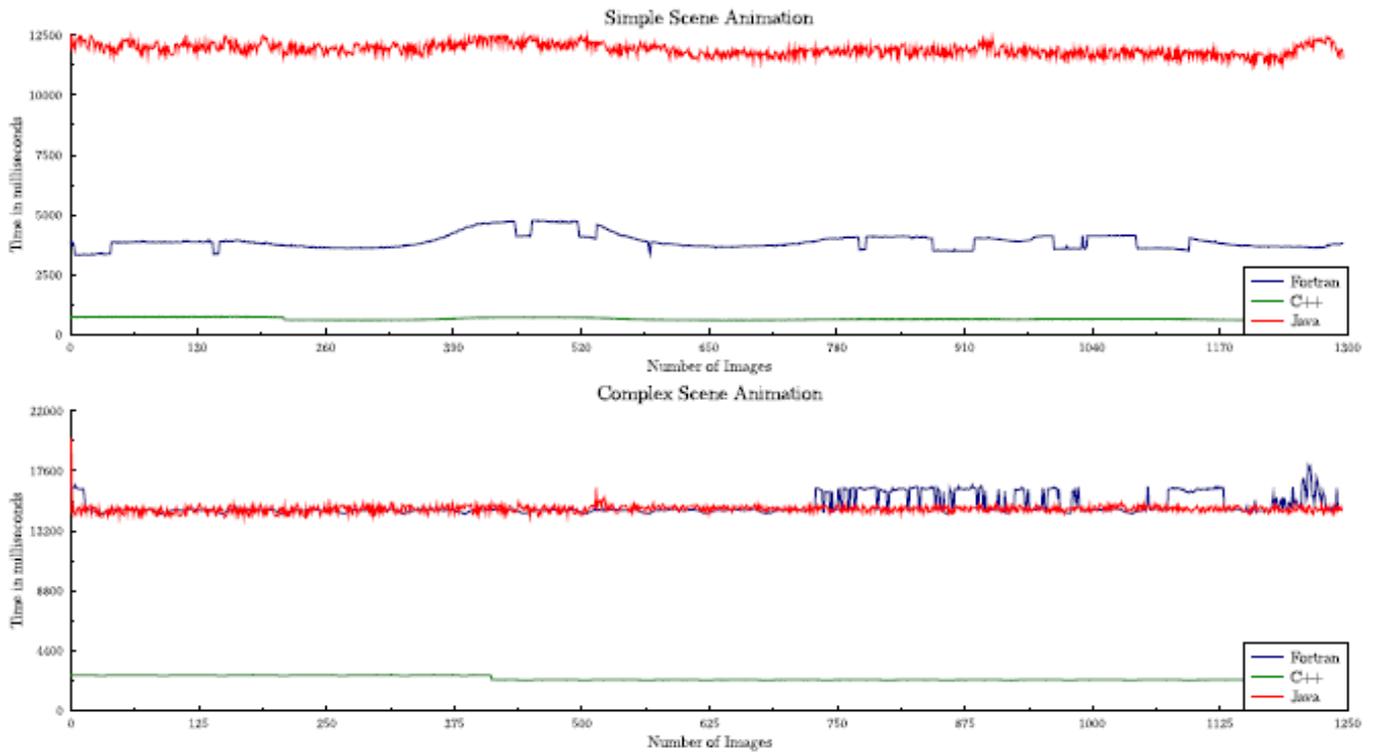

To simulate the real life environment where the load is always unbalanced we used **machines** that are continuously used for running different experiments and tests at IBM. That is why the results are non-uniform compared to Figure 5. **Compilers**: IBM XLF V13.1 [30], IBM XLC V 11.1 [30] and IBM J9 VM 1.6.0 [29] **Optimization** Flags: -O3 for XLF and XLC, and no flag used for Java.

**Figure 6: Runtime of the *Ray Tracing* Application (Using IBM XL Compilers), in Fortran, C++ and Java, for Rendering Images to Generate Animation.**

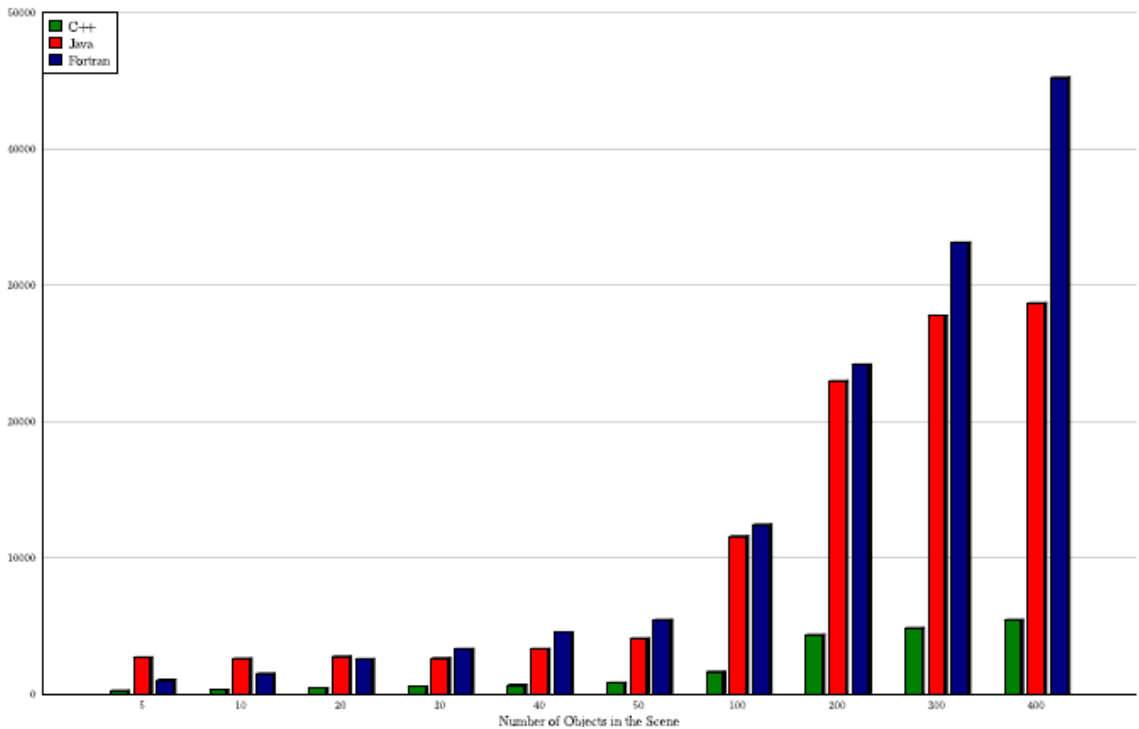

Following experimental setup was used to generate the data shown here:
Machine used: Power7 16 processors 4GB RAM; SUSE Linux Enterprise Server 11 SP1, 64 bit, Linux kernel: 2.6.32.12
Compilers: IBM XLF V13.1 [30], IBM XLC V 11.1 [30] and IBM J9 VM 1.6.0 [29]
Optimization Flags: -O3 for XLF and XLC, and no flag used for Java

To emphasize and compare the processing of the objects, the following simplified RayTrace function was used. We filtered the noise (File I/O, shading, reflection and refraction etc) from the CPU time. It was easy to make sure that this small part of the code is almost the same in all the three languages. To make the comparison more fair (specifically for Fortran), for iterating the objects in: Java we used an iterator to access the private objects; C++ we used an array of pointers to access the private objects; Fortran we directly access the public objects:

```
logical function RayTrace(color, ray, depth)
      hit = 0
      for i = 1, length(objects)
            if (hit(objects[i], ray)) then
                  object_hit = objects[i]
                  hit = 1
            end if
      end for
      material = get_material(object_hit)
      if (hit == 0) then
            return false
      end if
      color = get_color(material)
      return true
end function RayTrace
```

**Figure 7: Average Runtime of the Function *RenderOneImage()* of the *Ray Tracing* Application, Using Different Number of Objects (planes, spheres etc in the scene), in Fortran, C++ and Java; The Average is Computed over 30 Rendered Images.**

The top (Simple Scene Animation) and the bottom (Complex Scene Animation) animations contain 1296 and 1245 total number of images and, 5 and 56 total number of objects, respectively. The *ray tracing* application in Fortran, using the open source compiler gfortran shown in Figure 5, on average took 100% and 300% more time than the C++ application to render each image for the top and the bottom animations respectively. Java is the slowest when rendering animation for the Simple Scene but Fortran is 25% slower than Java when rendering the Complex Scene. Our first thought was that Fortran is spending more time in file I/O than the other two languages. But further instrumentation at the function level of the source code shows that in the three languages compared in this paper the *ray tracer* spends most of the time, almost 90%, in the recursive function *RayTrace()*. Due to the object oriented features of the *ray tracing* application, this difference shows that these features are not implemented as efficiently in the gfortran as in the g++ compiler. Here we also want to point out a fact that Java is slower at startup and is shown in our experimental results in Figure 5. As gfortran is an open source free compiler, and to make a precise conclusion, we also present and compare the results using the IBM XL compilers in Figure 6.

The *ray tracing* application in Fortran, using the IBM XLF compiler shown in Figure 6, on average took 400% and 500% more time than the C++ application to render each image for the top and the bottom animations respectively. Java is the slowest when rendering animation for the Simple Scene, but Fortran is only 30% faster than Java in SUSE Linux and 1.6% slower in Red Hat Linux when rendering the Complex Scene. These results confirm that our conclusion above also applies to the IBM XL compilers. We provide an assessment of these and other results in the next Section. Further analysis of the source code at the statement level shows that as we increase the number of objects in the scene the Fortran application took more time to render each image. Based on this analysis of object processing in the three languages, we present another comparison in Figure 7, which shows the average runtime of the function *RenderOneImage()* of the *ray tracing* application using a range of, from 5 - 400, objects. To emphasize and compare the processing of the objects in the scene we simplified the *RayTrace()* function as shown in Figure 7.

Figure 7 shows that Fortran remains ahead of Java when the number of objects are 20 or less. After 20 as we double the number of objects so does the runtime of Fortran processing these objects. With the introduction of object-oriented features and to make Fortran relevant, competitive and successful in this area both the open source and the commercial compilers need to improve this particular area. One of the reasons it is not optimized in these two popular compilers is that either the Fortran programmers (engineers and scientists) are not using or the use is still immature of the object-oriented features of the language. These object-oriented features are still very new to the Fortran language and will take some time to become mature and stable among programmers and the tool developers. We provide more insight into this in Section 6 As mentioned before this is one of the motivations of this paper to increase the use of these features among the programmers and improve the available tools.

**Table 3: Size Metrics for the *Ray Tracing* Application in Fortran, C++ and Java**

| Metric | Fortran | C++ | Java |
|---|---|---|---|
| Number of Files | 18 | 31 | 19 |
| Size in Bytes | 101554 | 74959 | 70965 |
| Size in Words | 11987 | 9287 | 8703 |
| Physical Lines | 3336 | 3254 | 2828 |

International Journal of Software Engineering & Applications (IJSEA), Vol.5, No.3, May 2014

Fortran has a total 18 files, 16 of which contain class definitions and 2 that contain global parameters. These 2 files can also be called global header files. C++ has a total of 16 header files and therefore has more files than Fortran and Java. In Java the file name and the class defined in the file should have the same name. There are a total of 24 classes in the Java *ray tracer* as shown in the Size-Level Metrics in Table 1, of which there are 5 inner classes, therefore there are only 19 files in Java. We mentioned Fortran as a verbose programming language which is confirmed by the larger sizes of the *ray tracing* application in Fortran listed in Table 3. It took more time to implement the application in Fortran than Java and C++. Java implementation took the least time. The size of the code also follow the same order as shown in Table 3.

# 6. BINARY ANALYSIS AND PROFILING

Comparing the performance of programming languages is a non-trivial task. As mentioned previously the purpose of the experiments and the results shown in the graphs and the tables of this paper, is to compare the performance of these object-oriented languages for efficiently handling and processing the objects. In this section we highlight this particular area of the Fortran language and argue that it needs improvement by profiling the runtime and analyzing the assembly codes generated by the three compilers. We use GNU compilers for C++ and Fortran and Oracle compiler for Java. We profiled a simple object-oriented application, implemented in the three languages, using the PIN tool [35]. PIN is a dynamic binary instrumentation tool, i.e: it performs instrumentation at runtime. We wrote a small PIN application to count the number of times a routine is executed (number of calls). Since the instrumentation is done at the binary level, it includes routines both from the application and from the dynamic libraries linked at the runtime. We also counted the total number of instructions executed in each of these routines. After examining the results we found that the Fortran application spent most of its time in the function (routine) *Sphere::SetPosition* as shown in Table 4. We also list the generated assembly and byte code for the function *Sphere::SetPosition* in Table 5.

Table 4: PIN Results (C++ and Fortran) for the Number of Calls and the Number of Instructions for the two Functions from the Simple Object Oriented Application (implemented in the three languages) are Shown for one Iteration.

| Language | Function | Image[1] | Address | Calls | Instructions |
|---|---|---|---|---|---|
| C++ | Sphere::SetPosition | simple | 0x400A90 | 393216000[2] | 1572864000 |
| Fortran | Sphere::SetPosition | simple | 0x400950 | 393216000[2] | 3145728000 |
| C++ | main | simple | 0x400830 | 1 | 5910828522 |
| Fortran | main | simple | 0x400970 | 1 | 9057340465 |

1  Image to which the function (routine) belongs, and can be the application or the library. Here it is the application.
2  These results are also confirmed by computing the number of calls to function *Sphere::SetPosition* from code
   of the simple object oriented application (implemented in the three languages): 1024 x 768 x 500 = 393216000.

The number of byte code instructions generated by the Java compiler as shown in Table 5 does not have any relation with the actual number of instructions executed. After profiling the Java application with Oracle JRockit Mission Control [22] we were able to determine that the function *Sphere::SetPosition* has been optimized by the runtime HotSpot JIT compiler but it was not clear what optimizations were performed. We also compiled the Java application into native code using the gcj [25] compiler version 4.4. gcj and g++ compilers use the same back-end, and the assembly code generated for the function *Sphere::SetPosition* by the gcj compiler was exactly the same as generated by the g++ compiler. Therefore in the next paragraph we discuss only, some of the differences in the generated assembly code, by the Fortran and the C++ compilers.

**Table 5: Generated Assembly (C++ and Fortran) and Byte Code (Java) for the Function Sphere::SetPosition Implemented as Part of the Simple Object Oriented Application (implemented in the three languages). Compilers used were g++, gfortran version 4.6 and Oracle javac version 1.6 on Linux 2.6.32 installed on Intel Core 2 Duo 64 bit machine. Optimization level O3 was used by the g++ and the gfortran compilers.**

| Language | Generated Assembly / Byte Code |
| --- | --- |
| C++[1] | ```
1  0x400a90 : f 2 0 f 11 47 08 movsd %xmm0, 0 x8(%r d i )
2  0x400a95 : f 2 0 f 11 4 f 10 movsd %xmm1, 0 x10(%r d i )
3  0x400a9a : f 2 0 f 11 57 18 movsd %xmm2, 0 x18(%r d i )
4  0x400a9f : c3 r e tq
``` |
| Fortran | ```
1  0x400950 : 48 8b 07 mov (%r d i ) ,%rax
2  0x400953 : 48 8b 36 mov (%r s i ) ,% r s i
3  0x400956 : 48 89 30 mov %r s i ,(%rax )
4  0x400959 : 48 8b 12 mov (%rdx ) ,%rdx
5  0x40095c : 48 89 10 mov %rdx ,(%rax )
6  0x40095f : 48 8b 11 mov (%rcx ) ,%rdx
7  0x400962 : 48 89 10 mov %rdx ,(%rax )
8  0x400965 : c3 r e tq
``` |
| Java | ```
1   0: aload_0 // Load parameter ( r e f e r e n c e ) onto the s tack
2   1: dload_1 // Load l o c a l v a r i a b l e onto the s tack
3   2: p u t f i e l d #2; // Set f i e l d to value i . e : x = X
4   5: aload_0
5   6: dload_3
6   7: p u t f i e l d #3; // Set f i e l d to value i . e : y = Y
7  10: aload_0
8  11: dload_5
9  13: p u t f i e l d #4; // Set f i e l d to value i . e : z = Z
10 16: r e turn
``` |

[1] All instructions can execute independently.
Instruction MOVSD %xmm0,0x8(%rdi) moves the data (double precision floating point) from register xmm0 to the memory location pointed to by the value stored at register rdi plus 8. MOVSD is a x86 SIMD instruction and xmm0 is one of the 128 bit registers added as Streaming SIMD extension (SSE) reference: Section 2.2.7 of [19]. Instruction MOV (%rdi),%rax moves the value stored at memory location pointed to by the value in register rdi to register rax [19].

The number of instructions are more than double in the assembly code generated by the Fortran compiler than the assembly code generated by the C++ compiler as shown in Table 5. The results of profiling in Table 4 also shows that the number of instructions executed by the Fortran function *Sphere:: SetPosition* are double than the C++ function *Sphere::SetPosition*. This gives one justification why Fortran is slower than C++ and Java. We can argue that the difference in number of instructions may not be important because almost all processors now a days have pipelines [26] and can execute instructions out-of-order [26]. Carefully examining the instructions generated by the C++ compiler reveals that the function have been optimized. It uses SIMD registers (see note to Table 5) and the instructions are not dependent [3] on each other, and therefore can be pipelined [3]. Whereas the instructions generated by the Fortran compiler use general registers and have multiple dependencies. Therefore these instructions cannot be efficiently scheduled [3] for pipelining. We will not go into the details why the Fortran compiler is not able to optimize this part of the code because it is out of the scope of this paper, but we point out some directions for future research to improve the object-oriented Fortran compilers. In object-oriented languages the selection of a target function in a dynamic dispatch [33] is very important and is only known at runtime [28, 24]. The compiler knows the abstract type of the object but not the concrete type. In our example the function *Shape::SetPosition* is abstract and is known to the compiler at compile time. The concrete type



International Journal of Software Engineering & Applications (IJSEA), Vol.5, No.3, May 2014

is the function *Sphere::SetPosition* where it is implemented and is only known at runtime. Therefore it can indirectly cause a compiler (which lacks the required analysis to get the information) to produce a poorly optimized code as in this case with the Fortran compiler. Every modern compiler [3] has a front-end and a back-end. The same back-end can be used for different languages that have different front-ends. Another argument that we make here is the use of the same back-end by g++ and gfortran compilers. Our profiling and generated assembly code analysis confirms that this is the case. Both the applications, in C++ and Fortran, are linked with the same libraries, except few, such as *cpu_time*, *set_args*, *set_options* and *os_error* etc. Most of the assembly code generated by both the compilers is similar but with some major differences. One of them is mentioned above and listed in Table 5. The other major difference is the number of instructions generated for the main function. The g++ compiler generated 88 instructions whereas the gfortran compiler generated 122 instructions for the main function. This difference is also evident from the number of instructions executed in the main function by C++ and Fortran applications as shown in Table 4. The number of assembly instructions generated for the following two very similar loops in the main function are: by gfortran 42 and by g++ 28. This loop is executed 393216000 times as shown in Table 4. This explains why the number of instructions executed by the Fortran main function are almost double than the number of instructions executed by the C++ main function, as shown in Table 4.

```
call cpu_time (time_start)
do n1 = 1, NUM_SPHERES
   COUNT = COUNT + 1
   n = n1
   call sphere_t(n1)%SetPosition(n,n+1.0,n+2.0)
end do
COUNT = COUNT / 100
call cpu_time (time_end)

            (a) Fortran
```

```
clock_t time_start = clock();
for (int n1 = 0; n1 < NUM_SPHERES; n1++) {
   COUNT = COUNT + 1;
   n = n1;
   sphere_t[n1].SetPosition(n,n+1.0,n+2.0);
}
COUNT = COUNT / 100.0;
clock_t time_end = clock();

            (b) C++
```

This part of the code is exactly where the application is making a virtual function call *Sphere::SetPosition*. This analysis further confirms that Fortran compiler is not optimizing the virtual function call overheads. One explanation for this is: that the front-end of the Fortran compiler is not communicating enough information (or similar information as the C++ compiler) to the back-end, which is required by the compiler to optimize the generated code for object handling and processing.

Based on the discussion in the above paragraphs we provide some pointers, for further exploration, and list some of the techniques that can improve such a code, as follows: (1) A better interprocedural analysis [3] such as complete information about the inheritance graph and the methods defined in each class. (2) Optimizations at link time, i.e: machine code optimization where more information is available about dynamic link libraries. (3) An improved front-end which collects and communicates enough information to the back-end for optimization. (4) Devirtualization [32]. For further information on optimizations for object-oriented languages the reader is referred to [32, 4, 1, 28, 27]. We have provided an insight into the simple object-oriented application implemented in the three languages in the hope that this will motivate language implementers and compiler developers to improve Fortran object handling and processing, and hence make it's use more prolific and general.

## 7. CONCLUSION

We have presented a comparative study to evaluate and compare Fortran with the two most popular languages Java and C++. Based on our literature survey this is the first study carried out to compare these languages by applying software metrics to an object-oriented application and comparing these results with the similarities and differences found in practice. This paper makes the following contributions:





1. An object-oriented application, a basic *ray tracer*, is implemented in Fortran, Java and C++. The *ray tracing* application is neither complex nor simple but complete enough to compare the object-oriented features of these languages. By using only one program we ensured there was only one set of requirements thus making the comparison homogeneous.
2. We applied software metrics to the *ray tracing* application and have highlighted some of the differences and similarities in these languages that are found in practice, like: Templates, object constructors and destructors, abstract data types, dynamic binding and reuse. The graphs and the tables shown highlight some of the differences and similarities in the three languages.
3. We have provided an insight into the binary analysis and profiling of a simple object-oriented application implemented in the three languages to highlight some of the inefficiencies present in the Fortran compiler, hoping to motivate language implementers and compiler developers to improve Fortran object handling and processing, and hence make it's use more prolific and general.
4. This study facilitates and encourages the reader to further explore, study and use these languages more effectively and productively especially Fortran.

Some of the important differences and findings in the three languages that are explored in this paper are:

1. Multiple inheritance and templates: Fortran does not support multiple inheritance and templates for the same reason they are not supported in Java as explained in Section 4.3.
2. Garbage collection: In Fortran and C++ deallocation of the objects is the responsibility of the programmer. In Java deallocation of the objects is taken care of by the language. Unlike Java the memory consumption of a Fortran and a C++ application can be fine tuned by an experienced programmer.
3. Pure object-oriented language: Java is a pure object-oriented language. Everything is an object in Java. In Fortran and C++ the programmer is not bound to use objects.
4. Object handling and processing: Fortran object handling and processing is not optimized in the two popular tools (GNU and IBM compilers) as shown in Sections 5 and 6. One of the reasons it is not optimized in these two popular compilers, as mentioned in Section 5, is that either the Fortran programmers are not using or the use is still immature of the object-oriented features of the language. This is one of the motivations of this paper to increase the use of this feature among programmers and improve the available tools.
5. Development time and cost: C++ and Java are more concise and clear but Fortran is more verbose and explicit. This in general can increase the development time and cost of software in Fortran compare to C++ and Java. This needs to be confirmed with more studies as mentioned below.

There are more similarities than differences as shown in Section 4 in the three languages compared. Therefore our study concludes that the object-oriented features (extensions) introduced in Fortran 2003 and 2008 are comparable to C++ and Java. There are some features that are missing or lacking in Fortran, such as support for anonymous classes, debugging, exception handling and string processing as part of the language. There are some non-object-oriented features in Fortran like the support of complex numbers, FORALL loop construct and parallel processing (we did not include this feature of Fortran in this study but interested readers can read more about this in [44, 41]) that are not available in C++ and Java as part of the language. We believe that with these extensions Fortran can be more productive and effective and be used as a popular standardized modern object-oriented parallel programming language in this multicore and the coming manycore era. In the future the *ray tracer* will be updated to





compare the parallel languages, such as Co-array Fortran [44, 41], UPC [18] and X10 [46]. We would also like to carry out a study on the programmer's productivity and efficiency in these three languages. Such as, how persons with different skill levels implement the same application and how easy it is to learn and program, these three languages.

**Author**

Shahid Alam is currently a PhD student in the Computer Science Department at University of Victoria, BC. He received his MASc degree from Carleton University, Ottawa, ON, in 2007. He has more than 5 years of experience working in the software industry. His research interests include programming languages, compilers, software engineering and binary analysis for software security. Currently he is looking into applying compiler, binary analysis and artificial intelligence techniques to automate and optimize malware analysis and detection.

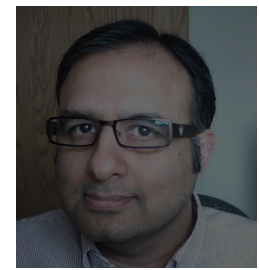